\documentclass[a4paper]{jpconf}
\usepackage{graphicx}
\begin{document}

\begin{flushright}
KOBE-TH-14-10
\end{flushright}

\title{Anisotropic Power-law Inflation: \\A counter example to the cosmic no-hair conjecture  }

\author{Jiro Soda}

\address{Department of Physics, Kobe University, Kobe 657-8501, Japan}

\ead{jiro@phys.sci.kobe-u.ac.jp}

\begin{abstract}
It is widely believed that anisotropy in the expansion of the universe 
will decay exponentially fast during inflation. This is often referred to as
the cosmic no-hair conjecture. 
However, we find a counter example to the cosmic no-hair conjecture in the context of supergravity. 
As a demonstration, we present an exact anisotropic power-law inflationary solution which is an 
attractor in the phase space. We emphasize that anisotropic inflation is quite generic
in the presence of anisotropic sources which couple with an inflaton.
\end{abstract}

\section{Introduction}
 
As is well known, the event horizon of black holes hides the initial conditions of the collapsed matter other than mass, charge, and angular momenta, which is named the black hole no-hair theorem~\cite{Israel:1967wq}. 
The similar thing should happen for inflation because of the cosmological event horizon, namely,
any initial conditions should go away beyond the cosmological event horizon. 
In fact, in the presence of the cosmological constant, there is a cosmic no-hair theorem proved by Wald~\cite{Wald}.
Even for a general accelerating universe driven by a scalar field, it is legitimate to expect that the anisotropy 
decays exponentially first in the presence of the cosmological event horizon. 
This prejudice is often referred to as the cosmic no-hair conjecture.   

 Historically, there have been challenges to the cosmic no-hair 
 conjecture~\cite{Ford:1989me}.  Unfortunately, it turned out that these models 
suffer from either the instability, or a fine tuning problem, or a naturalness problem~\cite{Himmetoglu:2008zp}. 
Recently, however, we have succeeded in finding stable anisotropic inflationary solutions in the context of supergravity. More precisely, 
we have shown that, in the presence of a gauge field coupled with an inflaton,
there could be small anisotropy in the expansion rate which never
decays during inflation. Since  anisotropic inflation is an attractor,
this can be regarded as a counter example to the cosmic no-hair conjecture~\cite{Watanabe,Kanno}.
Moreover, primordial fluctuations created quantum mechanically
during inflation also exhibit statistical anisotropy~\cite{Gum}.
Indeed, from the point of view of precision cosmology,
it is important to explore the role of gauge fields
in inflation~\cite{review}. 

It is well known that the supergravity models can be constrained
 by comparing predictions of inflation with cosmological observations.
 For example, the tilt of the spectrum gives interesting information
of the superpotential $W(\phi^i)$ and the Kaler potential $K(\phi^i,\bar{\phi}^i)$
which are functionals of complex scalar fields. Here, a bar represents
a complex conjugate. 
More concretely, the bosonic part of the action of supergravity reads
\begin{eqnarray}
S = \int d^4 x \sqrt{-g}
\left[ \frac{M_p^2}{2}R 
+ K_{i\bar{j}} \partial^\mu \phi^i \partial_\mu \bar{\phi}^{\bar{j}} 
   - e^K K^{i\bar{j}} D_i W D_{\bar{j}} \bar{W} 
   - f_{ab}(\phi) F^a_{\mu\nu} F^{b\mu\nu} \right]
\end{eqnarray}
where we have defined Kaler metric
$K_{i\bar{j}} = \partial^2 K/\partial \phi^i \partial \bar{\phi}^j$
and a covariant derivative  $D_i W=\partial W/\partial \phi^i + \partial K/\partial \phi^i W$. 
Here, $M_p $ represents the reduced Planck mass.
We note that the gauge kinetic function $f_{ab}(\phi)$ in front of the gauge
kinetic term $F^a_{\mu\nu} F^{b\mu\nu}$ 
could be nontrivial functions of the scalar fields.
Curiously, so far, the gauge field part has been neglected in most of 
discussion of inflation. 
The reason is partially due to the cosmic no-hair conjecture
which states that the anisotropy, curvature, and any matter will be diluted
once inflation commences. However, as we will see soon, this is not true when we look at percent
 level fine structures of inflationary scenarios. 
We find that gauge fields play an important role in  the early universe.
In fact, we will show that there exists anisotropic inflation and
the degree of the anisotropy we found is smaller than the slow roll parameter. 
Apparently, the expansion is almost isotropic. Nevertheless, anisotropic inflation
provides a qualitatively new picture of the universe. 

In this paper, as a demonstration, we consider the cases
where the functional form of the potential and the gauge kinetic functions
has an exponential dependence on the inflaton. In particular, 
we present exact anisotropic inflationary solutions.
Moreover, we argue anisotropic inflation becomes an attractor in the phase space. 

The organization of the paper is as follows.
In section 2, we introduce inflationary models where the gauge
field couples with an inflaton and obtain exact solutions which contain
anisotropic power-law inflationary solutions. 
In section 3, we show that the anisotropic inflation
is an attractor in the phase space. Thus, generic trajectories converge to 
anisotropic inflation. This implies that the cosmic no-hair conjecture does not hold
in general.
 The final section is devoted to the conclusion.

\section{Exact Anisotropic Power-law Inflationary Solutions}
\label{sc:basic}

In this section, we consider a simple model with exponential
potential and gauge kinetic functions and then find exact power-law inflationary solutions.
In addition to a well known isotropic power-law solution, we find 
an anisotropic power-law inflationary solution. 

We consider the following action for the metric $g_{\mu\nu}$, the inflaton
 field $\phi$ and the
gauge field $A_\mu$ coupled with $\phi$:
\begin{eqnarray}
S=\int d^4x\sqrt{-g}\left[~\frac{M_p^2}{2}R
-\frac{1}{2}\left(\partial_\mu\phi\right)\left(\partial^{\mu}\phi\right)
-V(\phi)-\frac{1}{4} f^2 (\phi) F_{\mu\nu}F^{\mu\nu}  
~\right] \ ,
\label{action1}
\end{eqnarray}
where $g$ is the determinant of the metric, $R$ is the
Ricci scalar, respectively.
 The field strength of the gauge field is defined by 
$F_{\mu\nu}=\partial_\mu A_\nu -\partial_\nu A_\mu$. 
Motivated by the dimensional reduction of
higher dimensional theory such as string theory, we assume the exponential potential
and the exponential gauge kinetic functions
\begin{eqnarray}
V (\phi) = V_0 e^{\lambda \frac{\phi}{ M_p}}  \ ,
\qquad
f(\phi) = f_0 e^{\rho \frac{\phi}{M_p} } \ .
\end{eqnarray}
In principle, the parameters $V_0 , f_0 , \lambda$, and $\rho$ 
can be determined once the compactification scheme is specified.
However, we leave those free hereafter. 

Using the gauge invariance, we can choose the gauge $A_0 =0$.
Without loosing the generality, we can take the $x$-axis in the direction
of the gauge field. Hence, we have the homogeneous fields of the form
$
A_\mu=(~0,~v(t),~0,~0~)
$
and 
$ 
\phi=\phi(t) \ .
$
As there exists the rotational symmetry in the $y$-$z$ plane,
we take the metric to be 
\begin{eqnarray}
ds^2=- dt^2+e^{2\alpha(t)}\left[~ 
e^{-4\sigma(t)}dx^2    
+e^{2\sigma(t)}\left( dy^2 + dz^2\right)~\right] \ ,
\label{metric}
\end{eqnarray}
where the cosmic time $t$ is used.
Here, $e^\alpha$ is an isotropic scale factor and $\sigma$ represents
a deviation from the isotropy.  It is easy to solve the equation for the gauge field as
\begin{eqnarray}
\dot{v} = f^{-2}(\phi ) e^{-\alpha -4\sigma}p_{A}, 
\label{eq:Ax}
\end{eqnarray}
where $p_A$ denotes a constant of integration. 
Using Eq.(\ref{eq:Ax}), we obtain equations
\begin{eqnarray}
\dot{\alpha}^2  &=& \dot{\sigma}^2
+\frac{1}{3M_p^2}\left[ \frac{1}{2} \dot{\phi}^2+V(\phi)
+\frac{p_{A}^2}{2}f^{-2} (\phi) e^{-4\alpha-4\sigma }  \right] \ , 
\label{hamiltonian:0}\\
\ddot{\alpha} &=& -3\dot{\alpha}^2 + \frac{1}{M_p^2} V(\phi )
 +\frac{p_{A}^2}{6M_p^2}f^{-2}(\phi )e^{-4\alpha -4\sigma}, 
\label{evolution:alpha:0}\\
\ddot{\sigma} &=& -3\dot{\alpha}\dot{\sigma} 
+ \frac{p_{A}^2}{3M_p^2}f^{-2}(\phi )e^{-4\alpha -4\sigma} 
\label{eq:sigma:0}, \\
\ddot{\phi} &=& -3\dot{\alpha}\dot{\phi} -V'(\phi ) 
+ p_{A}^2 f^{-3}(\phi )f'(\phi ) e^{-4\alpha -4\sigma } 
\label{eq:phi:0} \ .
\end{eqnarray}
Here, an overdot and a prime denote the derivative with respect to the 
cosmic time $t$ and $\phi$, respectively.

In the absence of gauge fields, it is known that there exists the power-law inflationary
solution.
Therefore, let us first seek the power-law solutions by assuming
\begin{eqnarray}
  \alpha = \zeta \log t \ , \hspace{1cm}
  \sigma = \eta \log t \ , \hspace{1cm}
  \frac{\phi}{M_p} = \xi \log t + \phi_0 \ .
\end{eqnarray}
Apparently, for a trivial gauge field $p_A =0$,
we have the isotropic power-law solution
\begin{eqnarray}
  \zeta = \frac{2}{\lambda^2} \ , \hspace{1cm} \eta=0 \ , \hspace{1cm} 
  \xi = - \frac{2}{\lambda} \ , \hspace{1cm} 
  \frac{V_0}{M^2_p} e^{\lambda \phi_0} = \frac{2(6-\lambda^2)}{\lambda^4} \ .
  \label{iso-power}
\end{eqnarray}
In this case, we have the spacetime
\begin{eqnarray}
 ds^2 = -dt^2 + t^{4/\lambda^2} \left( dx^2 +dy^2 + dz^2 \right) \ .
 \label{isotropic}
\end{eqnarray}
Hence, we need $\lambda \ll 1$ for obtaining the accelerating expansion. 

Next, interestingly, we see that there exists the other non-trivial exact 
solution in spite of the existence of the no-hair theorem~
\cite{Wald}.
From the hamiltonian constraint equation (\ref{hamiltonian:0}),
we set two relations
\begin{eqnarray}
  \lambda \xi = -2  \ , \hspace{1cm} \rho \xi +2 \zeta + 2\eta =1
  \label{A}
\end{eqnarray}
to have the same time dependence for each term. 
The latter relation is necessary 
only in the non-trivial gauge field case, $p_A \neq 0$.
Then, for the amplitudes to balance, we need 
\begin{eqnarray}
  -\zeta^2 +\eta^2 +\frac{1}{6} \xi^2 + \frac{1}{3} u +\frac{1}{6} w =0 \ ,
  \label{B}
\end{eqnarray}
where we have defined
\begin{eqnarray}
   u = \frac{V_0}{M_p^2} e^{\lambda \phi_0} \  ,  \hspace{1cm} 
   w = \frac{p_A^2 }{M_p^2} f_0^{-2} e^{-2\rho \phi_0}\,.
\end{eqnarray}
The equation for the scale factor (\ref{evolution:alpha:0}) 
under Eq.~(\ref{A}) yields
\begin{eqnarray}
  -\zeta + 3\zeta^2 -u - \frac{1}{6} w =0 \ .
  \label{C}
\end{eqnarray}
Similarly, the equation for the anisotropy (\ref{eq:sigma:0}) gives 
\begin{eqnarray}
  -\eta + 3 \zeta\eta - \frac{1}{3} w =0 \ .
  \label{D}
\end{eqnarray}
Finally, from the equation for the scalar (\ref{eq:phi:0}),
we obtain
\begin{eqnarray}
  -\xi + 3\zeta \xi + \lambda u -\rho w = 0  \ .
  \label{E}
\end{eqnarray}
Using Eqs.~(\ref{A}),~(\ref{C}) and (\ref{D}), we can solve $u$ and $w$ as
\begin{eqnarray}
u = \frac{9}{2} \zeta^2 - \frac{9}{4} \zeta - \frac{3\rho}{2\lambda} \zeta
                 + \frac{1}{4} + \frac{\rho}{2\lambda}  \ , \hspace{1cm}
w = -9 \zeta^2 + \frac{15}{2} \zeta + \frac{9\rho}{\lambda} \zeta 
                 -\frac{3}{2} -\frac{3\rho}{\lambda} \,.
                 \label{uw}
\end{eqnarray}
Substituting these results into Eq.~(\ref{E}), we obtain
\begin{eqnarray}
  \left( 3\zeta-1 \right) \left[ 6 \lambda \left( \lambda + 2\rho \right)\zeta 
  - \left( \lambda^2 + 8\rho \lambda + 12 \rho^2 + 8 \right) \right] =0 \ .
\end{eqnarray}
In the case of $\zeta=1/3$, we have $u=w=0$. Hence, it is not our desired 
solution. Thus, we have to choose
\begin{eqnarray}
 \zeta = \frac{\lambda^2 + 8 \rho \lambda + 12 \rho^2 +8}{6\lambda (\lambda + 2\rho)} \ .
\end{eqnarray}
Substituting this result into Eq.~(\ref{C}), we obtain
\begin{eqnarray}
 \eta = \frac{\lambda^2 + 2\rho \lambda -4 }{3\lambda (\lambda + 2\rho)}\,. 
\end{eqnarray}
From Eq.~(\ref{A}), we have
\begin{eqnarray}
 \xi = - \frac{2}{\lambda}\,.
\end{eqnarray}
Finally, Eq.~(\ref{uw}) reduce to
\begin{eqnarray}
 u = \frac{(\rho \lambda + 2\rho^2 +2)(-\lambda^2 + 4\rho \lambda +12 \rho^2 +8)}
      {2\lambda^2 (\lambda +2\rho )^2 } \ ,
\ 
 w = \frac{(\lambda^2 + 2\rho \lambda -4)(-\lambda^2 + 4\rho \lambda +12 \rho^2 +8)}
      {2\lambda^2 (\lambda +2\rho )^2 } \ .
\end{eqnarray}
Note that Eq.~(\ref{B}) is automatically satisfied.

In order to have inflation, we need $\lambda \ll 1$.
For these cases, $u$ is always positive.
Since $w$ should be also positive, we have the condition
$
 \lambda^2 + 2\rho \lambda > 4 \ .
$
Hence, $\rho $ must be much larger than one. 
Now, the spacetime reads
\begin{eqnarray}
  ds^2 = -dt^2 + t^{2\zeta-4\eta} dx^2 + t^{2\zeta +2\eta} 
\left( dy^2 + dz^2 \right) \ .
  \label{anisotropic}
\end{eqnarray}
This describes the anisotropically inflating universe.
The average expansion rate is determined by $\zeta$ and the average slow roll parameter
is given by
\begin{eqnarray}
\epsilon \equiv -\frac{\dot{H}}{H^2}
= \frac{6\lambda (\lambda + 2\rho)}{\lambda^2 + 8 \rho \lambda + 12 \rho^2 +8}
\ ,
\label{epsilon}
\end{eqnarray} 
where we have defined $H=\dot{\alpha}$. 
In the limit  $\lambda \ll 1$ and $\rho \gg 1$, 
this reduces to $\epsilon = \lambda /\rho$.  
Now, the anisotropy is characterized by the ratio between the anisotropic expansion rate and 
the isotropic expansion rate
\begin{eqnarray}
  \frac{\Sigma}{H} \equiv \frac{\dot{\sigma}}{\dot{\alpha}}
  = \frac{2(\lambda^2 + 2\rho \lambda -4) }
  {\lambda^2 + 8 \rho \lambda + 12 \rho^2 +8} 
= \frac{1}{3}I\epsilon    \,,
\hspace{1cm}
I=\frac{\lambda^2 + 2\rho\lambda - 4}{\lambda^2 + 2\rho\lambda}\,.
 \ . 
\label{soverh}
\end{eqnarray}
Remarkably, the anisotropy is proportional to the slow roll parameter
and the parameter $I$ is smaller than 1. Hence, typically, the anisotropy is quite small. 
Although the anisotropy is always small, it persists during inflation. Therefore, its effect on the cosmological observables 
is not negligible.

\section{Anisotropic Inflation: A counter example to the cosmic non-hair conjecture}
\label{dynamical}

\begin{figure}[htbp]
\begin{center}
\includegraphics[width=125mm]{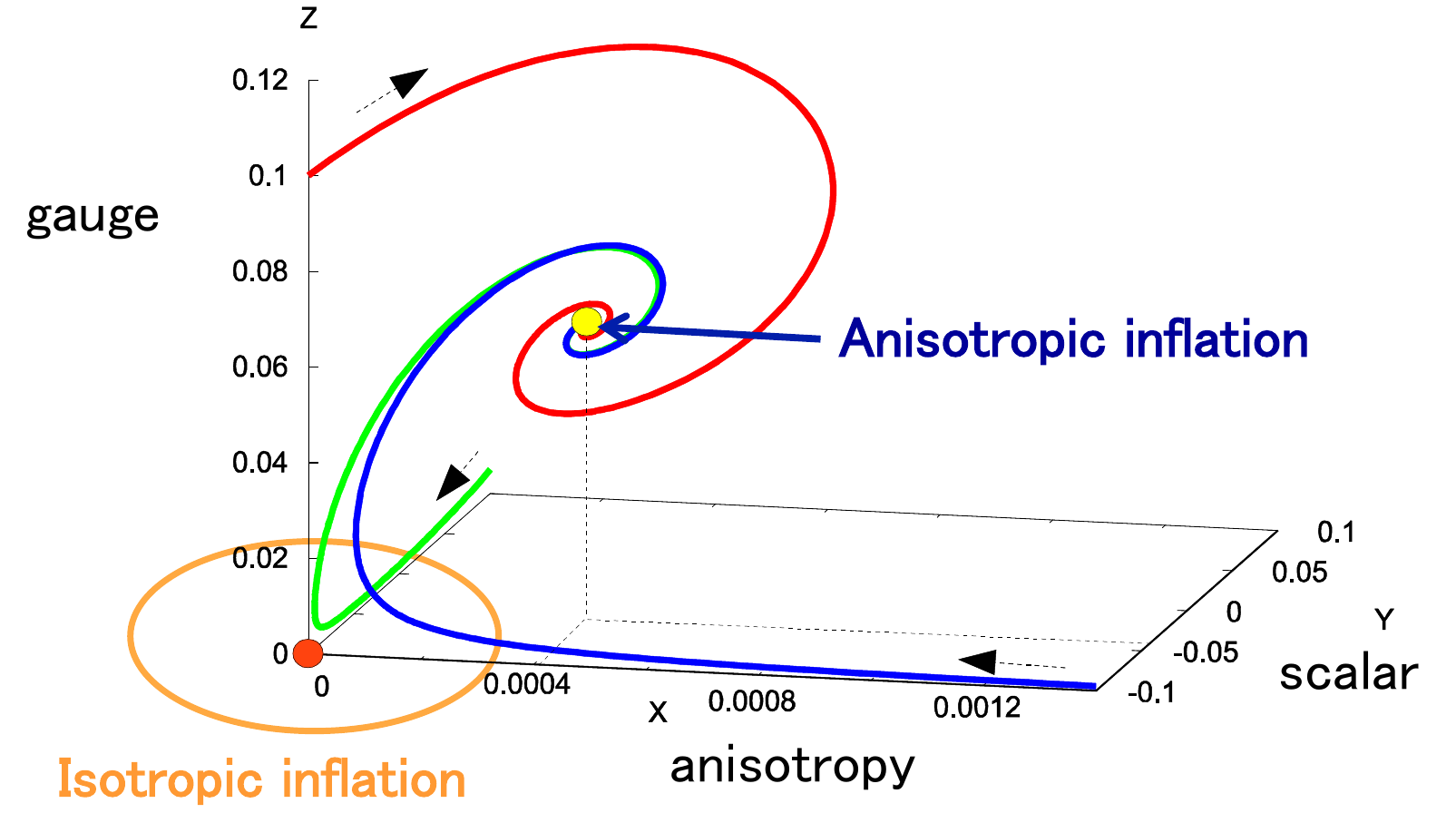}
\caption{The phase flow in the phase space is depicted for $\lambda = 0.1, \rho=50$. 
The axis $Z$ represents the energy density of the gauge field, the axis $X$ describes the anisotropy of the universe, and the axis $Y$ denotes the velocity of the inflaton.
The yellow and orange circles indicate the anisotropic and isotropic fixed points respectively. The trajectories converge to the anisotropic fixed point.} 
   \label{fg:phase}
\end{center}
\end{figure}

Now, we know two exact solutions, isotropic and anisotropic power-law inflationary solutions.
These solutions are not general solutions but particular solutions.
In order to eamine if the cosmic no-hair conjecture holds or not, we need to know the fate of generic solutions.
To this end, the dynamical analysis is useful.
Since two exact solutions correspond to two fixed points in the phase space, 
we did the stability analysis around the fixed points~\cite{Watanabe}.
It turns out that the isotropic fixed point indicated by orange circle in Fig.\ref{fg:phase} is a saddle point
 which is an attractor only on the two-dimensional $Z=0$ plane.
Thus, in the absence of gauge fileds, isotropic inflation is an attractor. Namely,
the cosmic no-hair holds in this special case.
On the other hand, we found that
the anisotropic fixed point is stable.
Thus, the end point of trajectories around the unstable isotropic power-law inflation
must be anisotropic power-law inflation.
In Fig.\ref{fg:phase}, we depicted the phase flow in the phase space for $\lambda = 0.1,\ \rho=50$.
We see that the trajectories converge to the anisotropic fixed point indicated by yellow circle.

When the initial conditions are close to the $Z=0$ plane, the trajectory approaches to the saddle point, but eventually it goes into
the attractor.
Interestingly, even if we start from isotropic universe, the final state of the universe is anisotropic (green line).
This can be regarded as the spontaneous breakdown of the rotational symmetry.

As we have shown, clearly, anisotropic inflation gives rise to a counter example to the cosmic no-hair 
conjecture. This feature is not specific to the exponential potential.
Although, in general cases, we do not have exact solutions, qualitative feature is almost the same.
Indeed, there are many concrete examples which violate the cosmic no-hair conjecture~\cite{Ward}.

We should note that the cosmological constant is assumed in
the cosmic no-hair theorem proved by Wald~\cite{Wald}. 
In the case of inflation, the inflaton can mimic the 
cosmological constant. Hence, the comic no-hair holds in the conventional cases.
However, in the presence of a non-trivial coupling between the inflaton
and the gauge field, the cosmic no-hair theorem cannot
be applicable anymore~\cite{Maleknejad:2012as}.

\section{Conclusion}

We have examined a supergravity model with a gauge kinetic function.
It turned out that exact anisotropic power-law inflationary solutions exist
when both the potential function for an inflaton and the gauge kinetic function
are exponential type. We showed that the degree of the anisotropy is 
proportional to the slow roll parameter. The slow roll parameter depends both on
the potential function and the gauge kinetic function. 
The dynamical system analysis told us that the anisotropic power-law inflation
 is an attractor. Therefore, the result we have mentioned in this paper 
presents a clear counter example to the cosmic no-hair conjecture.

The imprint of anisotropic inflation
can be found in the cosmic microwave background radiation~\cite{Gum}.
We should note that the statistical anisotropy could be large even when
the anisotropy in the expansion is quite small.
The result in this paper also has implication in cosmological magnetic
 fields~\cite{Kanno}.

\subsection{Acknowledgments}
I would like to thank Sugumi Kanno and Masa-aki Watanabe for collaborations.
This work was supported in part by the Grants-in-Aid for Scientific Research (C) No.25400251 and Grants-in-Aid for Scientific Research on Innovative Areas No.26104708.

\end{document}